\def\year{2019}\relax
\documentclass[letterpaper]{article} %DO NOT CHANGE THIS
\usepackage{aaai19}  %Required

\usepackage{times}  %Required
\usepackage{helvet}  %Required
\usepackage{courier}  %Required
\usepackage{url}  %Required
\usepackage{graphicx}  %Required
\usepackage{amsmath,amssymb,amsthm}
\usepackage{stmaryrd}  

\newtheorem{defn}{Definition}
\newtheorem{prop}[defn]{Proposition}
\newtheorem{theo}[defn]{Theorem}
\newtheorem{corolary}[defn]{Corollary}
\newtheorem{fact}[defn]{Fact}

\newcommand{\citeonline}[1]{\citeauthor{#1}~(\citeyear{#1})}
\newcommand{\citey}[1]{(\citeyear{#1})}
\begin{document}
% \frenchspacing
\title{Iterated Belief Base Revision: A Dynamic Epistemic Logic Approach}
\author{Marlo Souza\\
Department of Computer Science, UFBA \\
Salvador, Brazil\\
msouza1@ufba.br\\
\And
\'Alvaro Moreira\\
Institute of Informatics, UFRGS\\
Porto Alegre, Brazil \\
alvaro.moreira@inf.ufrgs.br\\
\And
Renata Vieira\\
Polytechnic School, PUCRS \\
Porto Alegre, Brazil\\
renata.vieira@pucrs.br
}

\maketitle

\begin{abstract}
AGM's belief revision is one of the main paradigms in the study of belief change operations. In this context, belief bases (prioritised bases) have been largely used to specify the agent's belief state - whether representing the agent's `explicit beliefs' or as a computational model for her belief state. While the connection of iterated AGM-like operations and their encoding in dynamic epistemic logics have been studied before, few works considered how well-known postulates from iterated belief revision theory can be characterised by means of belief bases and their counterpart in a dynamic epistemic logic. This work investigates how priority graphs, a syntactic representation of preference relations deeply connected to prioritised bases, can be used to characterise belief change operators, focusing on well-known postulates of Iterated Belief Change. We provide syntactic representations of belief change operators in a dynamic context, as well as new negative results regarding the possibility of representing an iterated belief revision operation using transformations on priority graphs.
\end{abstract}

\noindent 

%%%%%%%%%%%%%%%%%%%%%%%%%%%
\section{Introduction}

One of the most influential models for belief change is the so-called AGM paradigm, named after the authors of the seminal work \cite{AGM}. Although the AGM's approach has brought profound developments for the problem of belief dynamics,  influencing areas such as Computer Science, Artificial Intelligence, and Philosophy, it has not been immune to criticism \cite{Hansson,rott2000two,darwiche}.

Particularly, \citeonline{Hansson} criticises the use of deductively closed sets of formulas in the AGM paradigm, providing examples for which not just the meaning, but also the structure of the beliefs explicitly held by an agent may influence the change. This author proceeds to construct a different notion of belief revision, which became known as Belief Base Change, relying on the structure of the information believed by the agent.

While many studies \cite{nebel,WIL94,williams1995iterated,rott:shift} propose belief change operators based on the syntactic structure of an agent's explicit beliefs, or on the syntactic representation of an agent's belief state, few formal connections have been established between belief change postulates and these belief base change operators - some examples of works that investigate such connection are \cite{hansson1994kernel,ferme:JSL08,ferme:JLC17}. The axiomatic characterisation of syntactic-based belief change, however, has been concentrated on the one-shot behaviour of these operations and, thus, are not able to clarify their iterated or dynamic behaviour.

Belief base change operators are iterated, in the sense that they result in changing the belief state of the agent itself \cite{nayak2003dynamic}. As such, it is necessary to establish which formal properties these change operations satisfy on a dynamic sense, as studied in the literature of Iterated Belief Change \cite{darwiche,nayak2003dynamic,JIN06}.  

A step into connecting belief change operators and belief base change operators has been achieved by the work of Rott~\citey{rott:shift} and Liu~(\citeyear{liu2011reasoning}) which show that several belief change operations can be defined using transformations on some form of prioritised belief bases called priority graphs. Souza et al.~\citey{souzakr}, on the other hand, provide examples of belief change operations which cannot be defined in such a way.

While these latter works focus on using syntactic representations to construct belief change operations, they do not provide a strong connection between the syntactic-based transformations provided and the formal properties of the belief change operations they represent. To our knowledge,  there is no proposed characterisation in the literature of iterated belief change postulates based on syntactic representations of an agent's belief base. 

In this work, we study the relationship between iterated belief change postulates and properties of belief base change operators based on the counterpart of these operators as transformations on Liu's priority graphs (\citeyear{liu2011reasoning}). We obtain constraints on belief base change operators that guarantee satisfaction of some important postulates in the area of Iterated Belief Change. 

We also show the limits of expressibility of belief change operators by means of changes in belief bases.  These negative results provide us with a deeper understanding of the connections between belief base change and iterated belief change, helping to delineate the limits of the correspondence between semantic and syntactic theories of dynamic belief change.

This work is structured as follows: we start discussing the background knowledge in the next cection. Then,  we study how iterated belief change postulates can be characterised through transformations on priority graphs. Further, we show how our results can be used to construct syntactic representations of an iterated belief change operation, and in Section \textbf{Negative Results} we present some impossibility results for this characterisation.  Finally, we discuss the related literature and present some final considerations.

%%%%%%%%%%%%%%%%%%%%%%%%%%%%%%
\section{Background}

Much work has been conducted investigating syntactic representations of the explicit beliefs of an agent  \cite{Hansson,WIL94,rott:shift,liu:priority,baltag2016beliefs}. It is well-known that an agent's belief state - usually represented by a preference relation among worlds - can also be syntactically represented by means of orders among sentences, as well-investigated by \citeonline{lafage2005propositional}. 

We start by  introducing the syntactic representations we will use in this work to encode an agent's belief base. This structure, known as a priority graph, or P-graph for short,  was proposed by Liu~\citey{liu:priority,liu2011reasoning} and was further developed by Van Benthem, Grossi and Liu~\citey{liu:deontics}, and Souza et al~\citey{souzakr}. We will also present their relation to preference models - a generalisation of the models used in the area of belief revision to encode an agent's belief state. With this connection, we will be able to characterise well-known postulates of iterated belief revision using operations on priority graphs, thus connecting belief base change operations and iterated belief revision.

\begin{defn} \cite{liu2011reasoning}
Let $P$ be a countable set of propositional symbols and $\mathcal{L}_0(P)$ the language of classical propositional sentences over the set $P$. A P-graph is a tuple \mbox{$G = \langle \Phi, \prec \rangle$} where $\Phi \subset \mathcal{L}_0(P)$, is a set of propositional sentences and $\prec$ is a strict partial order on $\Phi$.
\end{defn}

Given a non-empty set of possible worlds $W$ and a valuation of propositional symbols over worlds of $W$, the order $\prec$ of a P-graph can be lifted to an order on worlds. Such an ordering can be called a preference  (or a plausibility) relation. 
  
\begin{defn}\cite{liu2011reasoning}
Let $G = \langle \Phi, \prec\rangle$ be a P-graph,  and ${v: P \rightarrow 2^W}$ be a valuation function of propositions over a non-empty set $W$ of possible worlds. The preference relation $\leq_G$ on $W$ induced by P-graph $G$ is defined as follows:
\[
\begin{array}{l}
w \leq_G w' ~~ \mathit{iff}  ~~~\forall \varphi \in \Phi: ((w' \vDash\varphi \Rightarrow w \vDash\varphi) ~ \mathit{or}\\
\quad\quad\quad\quad\quad\quad\quad  \exists \psi \in \Phi : \psi \prec \varphi, ~ w\vDash \psi,\mathit{and} ~w'\not\vDash\psi)
 \end{array}
 \]
\end{defn}

More yet, if a P-graph $G$ is finite, \citeonline{liu:deontics} have shown that a preference relation $\leq_G$, defined as above, satisfies well-foundedness - a property deeply connected with Lewis Limit Assumption, commonly required for semantic models of an agent's belief state.

It is worthy of notice that preference relations are a common semantic representation of an agents belief state. In fact, in Belief Revision Theory \cite{AGM}, an important semantic representation of agents belief states was proposed by Adam Grove, which is commonly known as Grove's systems of spheres (SOS) \cite{grove}, or simply as Grove's spheres.

A generalisation of Grove's spheres was given by Girard~\citey{girard2008modal}, called preference models or order models, in the context of Preference Logic. Similar models have been proposed before in the context of Qualitative Decision Theory \cite{boutilier:94}, Non-monotonic Reasoning \cite{kraus1990nonmonotonic}, among others. These models have been applied to the study of Dynamic Belief Revision by the work of Baltag and Smets~\cite{BAL08}, Girard and Rott~\citey{girard2014belief}, Liu~\citey{liu2011reasoning}, and Souza et al.~\citey{souzakr,souza:dali} and proved to be an expressible model for an agent's belief state.  In this work, we will adopt Souza's~\citey{souzaphd} definition of preference models - which requires the preference relation $\leq$ to have a well-founded strict part.

\begin{defn}\cite{souzaphd}
A preference model is a tuple $M = \langle W, \leq, v\rangle$ where $W$ is a set of possible worlds, $\leq$ is a reflexive, transitive relation over $W$ with a well-founded strict part, and $v: P \rightarrow 2^W$ a valuation function.
\end{defn}

From the above definitions, it is easy to see that from a P-graph we can construct a preference model by taking the preference relation induced by such a graph.

\begin{defn}\label{def:model}
Let $G = \langle \Phi, \prec\rangle$ be a P-graph and  let $M = \langle W, \leq, v\rangle$ be a preference model. We say $M$ is induced by $G$ iff $\leq \,=\, \leq_G$.
\end{defn}

The induction of preference models from P-graphs raises the question about the relations between these two structures. \citeonline{liu2011reasoning} shows that any finite model with a reflexive and transitive accessibility relation has an equivalent P-graph.

\begin{theo}\label{teo:pgraph}\cite{liu2011reasoning}
Let  $M = \langle W, R\rangle$ be a mono-modal Kripke structure. The following two statements are equivalent:
\begin{enumerate}
\item The relation $R$ is reflexive and transitive;
\item There is a priority graph $G=(\Phi,\prec)$ and a valuation $v$ s.t. $\forall w,w' \in W. ~w R w'$ iff $w \leq_G w'$.
\end{enumerate}
\end{theo}

It is well-known from work on Iterated Belief Revision~\cite{darwiche,nayak2003dynamic} that dynamic belief change operations can be described by a transformation in the agent's belief state. 

As such, we can define dynamic belief change operators using the following notion of dynamic operators on preference models,  where $\mathbb{M}(\mathcal{L}_\leq(P))$ is the class of all preference models for a logic language $\mathcal{L}_\leq(P)$ \cite{souza:dali}:

\begin{defn}\cite{souza:dali}
Let \mbox{$\star: \mathbb{M}(\mathcal{L}_\leq(P))\times \mathcal{L}_0(P) \rightarrow \mathbb{M}(\mathcal{L}_\leq(P))$}, we say $\star$ is a dynamic operator on preference models if for any preference model
$M= \langle W,\leq, v\rangle$ and formula 
$\varphi\in \mathcal{L}_0$, we have that $\star(M,\varphi) = \langle W, \leq_\star, v\rangle$. In other words, an operation on preference models is called a dynamic operator iff it only changes the relation of preference models.
\end{defn}

Liu et al~\citey{liu2011reasoning,liu:deontics} shows that these dynamic belief changes can also be described by means of changes in the priority graphs representing the agent's belief base. In the following,  $\mathbb{G}(P)$ denotes the set of all P-graphs constructed over a set $P$ of propositional symbols. 

\begin{defn}
We call a P-graph transformation any function $\dagger: \mathbb{G}(P)\times \mathcal{L}_0(P) \rightarrow \mathbb{G}(P)$.
\end{defn}

A P-graph transformation is, thus, a transformation in the agent's belief base, as represented by a priority graph. 

Since P-graphs and preference models are translatable into one another, it is easy to connect P-graph transformations and dynamic operators as well. 

\begin{defn}\label{def:induced}
Let \mbox{${\star: \mathbb{M}(\mathcal{L}_\leq(P))\times \mathcal{L}_0(P) \rightarrow \mathbb{M}(\mathcal{L}_\leq(P))}$} be a dynamic operator and ${\dagger: \mathbb{G}(P)\times \mathcal{L}_0(P) \rightarrow \mathbb{G}(P)}$ be a P-graph transformation. We say $\star$ is induced by $\dagger$ if for any preference model ${M\in \mathbb{M}(\mathcal{L}_\leq(P))}$ and any P-graph $G \in \mathbb{G}(P)$, if $M$ is induced by $G$ then the preference model $\star(M,\varphi)$ is induced by the \mbox{P-graph} $\dagger(G,\varphi)$, where  $\varphi$ is any propositional formula in $\mathcal{L}_0(P)$,
\end{defn}

Notice that not all P-graph transformations induce dynamic operators.  We say that syntactically different P-graphs are equivalent if they induce the same preference model. As such, if a P-graph transformation changes equivalent P-graphs in inconsistent ways, no dynamic operator can satisfy the condition of Definition~\ref{def:induced}.

For example, consider a P-graph transformation that changes the P-graph $p \prec q$ into the P-graph $p \prec q$, and changes the P-graph $p\wedge q \prec p \wedge \neg q \prec \neg p \wedge q \prec \neg p \wedge \neg q$ into $p\wedge q \prec \neg p \wedge q \prec p \wedge \neg q \prec \neg p \wedge \neg q$. Such a transformation cannot induce any dynamic operator since the original P-graphs are equivalent, i.e., induce the same models, but the resulting P-graphs are not. As such, we define the notion of relevant graph transformation.

\begin{defn}\label{def:relevant} 
We say that a be a P-graph transformation $\dagger$ is relevant if there is some dynamic operator $\star$ that is induced by it. 
\end{defn}

Notice that, the existence of relevant P-graph transformations is guaranteed by the previous representation results of \citeonline{liu2011reasoning} and \citeonline{liu:deontics} on the characterisation of some dynamic operators by means of changes in P-graphs. 

%%%%%%%%%%%%%%%%%%%%%%%%%%%%%%%%%
\section{Priority Changes Satisfying Postulates for Iterated Belief Changes}

In this work, we investigate the connection between the properties satisfied by  dynamic belief change operations, focusing on the postulates studied in the field of Iterated Belief Revision, and the properties satisfied by the P-graph transformations that encode these operations. We aim to  understand better which belief change operations can or cannot be encoded this way and, thus, the differences between dynamic belief change based on semantic models and based on syntactic representations.

As such, the main results of our work can be stated as the characterisations provided in Propositions~\ref{prop:DP1}, \ref{prop:DP2}, \ref{prop:DP3}, \ref{prop:DP4},  \ref{prop:Rec} and \ref{prop:Ind}, as well as the negative results provided in Fact~\ref{fact:CB} and Corollary~\ref{cor:CB}. An interesting application of these results will be obtained in the next Section, in which we construct a P-graph transformation to implement the operation of lexicographic revision \cite{nayak2003dynamic} and obtain, as a corollary, the harmony result proved by \citeonline{liu:deontics} stating the correctness of this transformation.

The use of postulates to encode rational constraints in the way an agent must change her beliefs is a defining characteristic of the AGM approach to Belief Revision \cite{AGM}. These postulates, however, are usually defined by means of constraints on changes in the agent's belief state, thus, in our case, on preference models. We must, then, define what it means for a P-graph transformation to satisfy some postulate (or property) for belief change operators.

\begin{defn}\label{def:postulate} 
We say that a P-graph transformation $\dagger$ satisfies a postulate $\mathcal{P}$ if (i) $\dagger$ is relevant and (ii) any dynamic operator $\star$ induced by $\dagger$ satisfies postulate $\mathcal{P}$. 
\end{defn}

AGM belief revision says very little about how to change one agent's beliefs repeatedly. In fact, it has been observed that the AGM approach allows some counter-intuitive behaviour in the iterated case \cite{darwiche}. As a result, different authors have proposed additional postulates that encode rational ways to change one's beliefs in an iterated way.

Most famous among them is the work of Darwiche and Pearl~\citey{darwiche}. They propose a set of postulates known as the DP postulates for iterated revision. Let $S= \langle W,\leq \rangle$ be a SOS and $S'= \langle W,\leq_{*\varphi} \rangle$ the result of revising the SOS $S$ by a formula $\varphi$, the DP postulates can be stated as:

\begin{enumerate}
\item[]\textbf{(\textsc{DP}-1)} If $w,w'\in \llbracket \varphi\rrbracket$, then $w\leq_{*\varphi} w'$ iff $w\leq w'$
\item[] \textbf{(\textsc{DP}-2)} If $w,w'\not \in \llbracket \varphi\rrbracket$, then $w\leq_{*\varphi} w'$ iff $w\leq w'$
\item[]\textbf{(\textsc{DP}-3)} If $w\! \in \! \llbracket \varphi\rrbracket$ and $w' \! \not \in \! \llbracket \varphi\rrbracket$, then  $w \! < \! w'$ $\Rightarrow$ $w <_{*\varphi} w'$ 
\item[] \textbf{(\textsc{DP}-4)} If $w\! \in \! \llbracket \varphi\rrbracket$ and $w' \! \not \in \! \llbracket \varphi\rrbracket$, then  $w \! \leq \! w'$ $\Rightarrow$ $w \leq_{*\varphi} w'$ 
\end{enumerate}

We want to provide a set of constraints on P-graph transformations that guarantee that the  dynamic operators induced by them satisfy these postulates. With that, we wish to study the connections between the classes of belief base revision operations and iterated belief change operators.

Let us start with \textsc{DP}-1. The postulate \textsc{DP}-1 states that for any two worlds $w,w'$ satisfying $\varphi$, there is no reason for their relative order to change in the agent's belief state after revision. In terms of P-graphs, this means that for any formula that $w'$ satisfies in the changed priority graph, either $w$ must also satisfy it, or there must be a formula that is preferred to it and that $w$ satisfies. We can ensure this property guaranteeing that the resulting P-graph is related to the original by a set of constraints in how it must be changed.

\begin{prop}\label{prop:DP1}
Let  $\dagger : \mathbb{G}(P)\times \mathcal{L}_0(P) \rightarrow \mathbb{G}(P)$ be a relevant P-graph transformation. If, for any P-graph $G = \langle \Phi, \prec\rangle$ and propositional formula $\varphi \in \mathcal{L}_0(P)$, the P-graph $\dagger(G,\varphi)= \langle \Phi_\dagger, \prec_\dagger\rangle$ satisfies the conditions below, then $\dagger$ satisfies \mbox{\textsc{DP-1}}:
\begin{enumerate}
\item For all $\xi \in \Phi$, there is some $\xi' \in \Phi_\dagger$ s.t.
\begin{enumerate} 
\item $\varphi \wedge \xi \equiv \varphi \wedge \xi'$ and
\item $\forall \psi' \in \Phi_\dagger$, if $\psi' \prec_\dagger \xi'$ then
\begin{enumerate}
\item[] $\psi'\equiv \varphi$ or
\item[] there is $\psi \in \Phi$ s.t. $ \varphi \wedge \psi \equiv  \varphi \wedge \psi'$ and $\psi\prec \xi$;
\end{enumerate}
\end{enumerate}
\item For all $\xi \in \Phi_\dagger$, $\xi \equiv \varphi$ or there is some $\xi' \in \Phi$ s.t.
\begin{enumerate}
\item  $ \varphi \wedge \xi \equiv  \varphi \wedge \xi'$ and  
\item   $\forall \psi' \in \Phi$, if $\psi' \prec \xi'$ then there is $\psi \in \Phi_\dagger$ s.t.
\begin{enumerate}
\item[] $\varphi \wedge \psi \equiv  \varphi \wedge \psi'$ and
\item[] $\psi\prec_\dagger \xi$.
\end{enumerate}
\end{enumerate}
\end{enumerate}
\end{prop}
\begin{proof}[Proof]
By Definition \ref{def:postulate} we have to prove that any dynamic operator $\star$ induced by $\dagger$ satisfies the postulate \textsc{DP}-1 .
Let $\dagger$ be a P-graph transformation satisfying the conditions above, $\star$ a dynamic operator induced by $\dagger$, \mbox{$M = \langle W, \leq, v\rangle$} a preference model, and \mbox{$\varphi \in \mathcal{L}_0$} a propositional formula. Also given \mbox{$M_{\star\varphi}=\star(M,\varphi) = \langle W, \leq_{\star\varphi},v\rangle$}, take a P-graph $G = \langle \Phi, \prec\rangle$ inducing $M$ s.t. \mbox{$\dagger(G,\varphi) = \langle \Phi_\dagger, \prec_\dagger \rangle$} induces $\star(M,\varphi)$.

$\Leftarrow$:\\
Take $w,w'\in \llbracket \varphi \rrbracket_M$ such that $w\leq_{\star \varphi}w'$, and take $\xi \in \Phi$ such that \mbox{$M,w'\vDash \xi$} - notice that the case in which no such formula exists is trivial, since necessarily $w\leq w'$ in such case. Then $M_{\star\varphi},w'\vDash \xi$, since $\xi$ is propositional formula. 

\noindent Since $\xi \in \Phi$ then, by condition 1(a), there is some $\xi'\in \Phi_\dagger$ s.t. $\varphi \wedge \xi \equiv \varphi \wedge \xi'$. As such, $M_{\star\varphi},w' \vDash \xi'$, given that $\star$ is induced by $\dagger$. As $w\leq_{\star \varphi}w'$, either:
\begin{itemize}
    \item[] (i) $M_{\star\varphi},w \vDash \xi'$, or
    \item[] (ii) there is some $\psi'\in \Phi_\dagger$ s.t. $\psi'\prec_\dagger \xi'$, $M_{\star\varphi},w \vDash \psi'$ and $M_{\star\varphi},w' \not\vDash \psi'$. Hence, by condition 1 (b), there is some $\psi \in \Phi$ s.t. $\varphi \wedge \psi \equiv \varphi \wedge \psi'$ and $\psi \prec \xi$. As such $M,w\vDash \psi$ and $M,w'\not\vDash \psi$. 
\end{itemize}
From (i) and (ii) we conclude that $w\leq w'$

$\Rightarrow$: Similar to the case before. Take $w,w'\in \llbracket \varphi \rrbracket_M$ s.t. \mbox{$w\leq w'$} and $\xi \in \Phi_\dagger$, use condition 2 to conclude that \mbox{$w\leq_{\star \varphi} w'$}.
\end{proof}

The postulate \textsc{DP}-2  describes the same information as \textsc{DP}-1, only restricted to those worlds that do not satisfy $\varphi$. As such, we can provide a similar characterisation.

\begin{prop}\label{prop:DP2}
Let  $\dagger : \mathbb{G}(P)\times \mathcal{L}_0(P) \rightarrow \mathbb{G}(P)$ be a relevant P-graph transformation. If, for any P-graph $G = \langle \Phi, \prec\rangle$ and propositional formula $\varphi \in \mathcal{L}_0(P)$, the P-graph $\dagger (G,\varphi)= \langle \Phi_\dagger, \prec_\dagger\rangle$ satisfies the conditions below, then $\dagger$ satisfies \mbox{\textsc{DP-2}} :
\begin{enumerate}
\item For all $\xi \in \Phi$, there is some $\xi' \in \Phi_\dagger$ s.t. 
\begin{enumerate}
\item $\neg\varphi \wedge \xi \equiv \neg\varphi \wedge \xi'$ and
\item $\forall \psi \in \Phi$, if $\psi \prec \xi$ then  there is $\psi' \in \Phi_\dagger$ s.t. 
\begin{enumerate}
\item $ \neg\varphi \wedge \psi \equiv  \neg\varphi \wedge \psi'$ and
\item $\psi'\prec_\dagger \xi'$;
\end{enumerate}
\end{enumerate}
\item For all $\xi \in \Phi_\dagger$, $\xi \equiv \varphi$ or there is some $\xi' \in \Phi$ s.t. 
\begin{enumerate}
\item $\neg\varphi \wedge \xi \equiv  \neg\varphi \wedge \xi'$ and 
\item $\forall \psi \in \Phi_\dagger$, if $\psi \prec_\dagger \xi$ then 
\begin{enumerate}
\item[] $\psi\equiv \varphi$ or
\item[] there is $\psi' \in \Phi$ s.t. $\neg\varphi \wedge \psi \equiv  \neg\varphi \wedge \psi'$ and $\psi'\prec \xi'$.
\end{enumerate}
\end{enumerate}
\end{enumerate}
\end{prop}
\begin{proof} Similar to that of Proposition~\ref{prop:DP1}.
\end{proof}

The postulate \textsc{DP}-3  states that for any two worlds $w$ satisfying $\varphi$ and $w'$ not satisfying it,  after revision by $\varphi$, if $w$ was preferable to $w'$ then it must continue to be so in the agent's belief state. In terms of P-graphs, we can guarantee this condition  requiring that, if there was a formula in the original P-graph that $w$ satisfied and $w'$ did not, there must be a formula in the revised P-graph s.t. $w$ satisfies and $w'$ does not. Therefore, we can characterise \textsc{DP}-3.
\begin{prop}\label{prop:DP3}
Let  $\dagger : \mathbb{G}(P)\times \mathcal{L}_0(P)\rightarrow \mathbb{G}(P)$ be a relevant P-graph transformation. If, for any P-graph $G = \langle \Phi, \prec\rangle$ and propositional formula $\varphi \in \mathcal{L}_0(P)$, the P-graph $\dagger(G,\varphi)= \langle \Phi_\dagger, \prec_\dagger\rangle$ satisfies the condition below, then $\dagger$ satisfies \mbox{\textsc{DP-3}} :
\begin{itemize}
\item For all $\xi \in \Phi$ or there is some $\xi' \in \Phi_\dagger$ s.t. 
\begin{itemize}
\item[(a)] $\varphi \wedge \xi \vdash  \xi'$, 
\item[(b)] $\neg\varphi \wedge \xi' \vdash  \xi$, and
\item[(c)] $\forall \psi' \in \Phi_\dagger$, if $\psi' \prec_\dagger \xi'$ then $\psi'\equiv \varphi$ or there is $\psi \in \Phi$ s.t. $\varphi \wedge \psi \vdash  \psi'$, $\neg\varphi \wedge \psi' \vdash  \psi$ and $\psi\prec \xi$.
\end{itemize}
\end{itemize}
\end{prop}
\begin{proof}[Proof]
Let $\dagger$ be a P-graph transformation satisfying the conditions above, $\star$ a dynamic operator induced by $\dagger$, $M = \langle W, \leq, v\rangle$ a preference model and $\varphi \in \mathcal{L}_0(P)$ a propositional formula. Also given $M_{\star\varphi}=\star(M,\varphi) = \langle W, \leq_{\star\varphi},v\rangle$, take $G = \langle \Phi, \prec\rangle$ be a P-graph inducing $M$ s.t. $\dagger(G,\varphi) = \langle \Phi_\dagger, \prec_\dagger \rangle$ induces $\star(M,\varphi)$.

Take $w,w'\in W$ s.t. $M,w\vDash \varphi$, $M,w'\not\vDash \varphi$ and $w<w'$. From $w<w'$ we conclude that there is some $\xi \in \Phi$ s.t. $M,w\vDash \xi$, $M,w'\not\vDash \xi$ and for all $\forall \psi \prec \xi$, if $M,w'\vDash \psi$ then $M,w\vDash \psi$. Since $\xi \in \Phi$, there is some $\xi' \in \Phi_\dagger$ s.t. $\varphi \wedge \xi \vdash  \xi'$, $\neg\varphi \wedge \xi' \vdash  \xi$. As such $M_{\star\varphi},w\vDash \xi'$,  $M_{\star\varphi},w'\not\vDash \xi$ and for all $\psi'\prec_\dagger \xi'$, if $M_{\star\varphi},w'\vDash \psi'$ then there is $\psi \in \Phi$ s.t. $\varphi \wedge \psi \vdash  \psi'$, $\neg\varphi \wedge \psi' \vdash  \psi$ and $\psi\prec \xi$. As such $M,w'\vDash \psi$ and, thus, $M,w\vDash \psi$. Since $\varphi \wedge \psi \vdash  \psi'$, then $M_{\star\varphi},w\vDash \psi'$, thus $w\prec_\dagger w'$.
\end{proof}

Along the same lines, as we did for \textsc{DP}-3, we can characterise postulate \textsc{DP}-4.

\begin{prop}\label{prop:DP4}
Let  $\dagger : \mathbb{G}(P)\times \mathcal{L}_0(P) \rightarrow \mathbb{G}(P)$ be a relevant P-graph transformation. If, for any P-graph $G = \langle \Phi, \prec\rangle$ and propositional formula $\varphi \in \mathcal{L}_0(P)$, the P-graph $\dagger(G,\varphi)= \langle \Phi_\dagger, \prec_\dagger\rangle$ satisfies the condition below, then $\dagger$ satisfies \mbox{\textsc{DP-4}} :
\begin{itemize}
\item For all $\xi \in \Phi_\dagger$, $\xi \equiv \varphi$ or there is some $\xi' \in \Phi$ s.t.
\begin{itemize}
\item[(a)] $\varphi \wedge \xi' \vdash  \xi$,
\item[(b)] $\neg\varphi \wedge \xi \vdash  \xi'$, and
\item[(c)] $\forall \psi' \in \Phi$, if $\psi' \prec \xi'$ then there is $\psi \in \Phi_\dagger$ s.t. $\varphi \wedge \psi' \vdash  \psi$, $\neg\varphi \wedge \psi \vdash  \psi'$ and $\psi\prec \xi$.
\end{itemize}
\end{itemize}
\end{prop}
\begin{proof}[Proof]
Let $\dagger$ be a P-graph transformation satisfying the conditions above, $\star$ a dynamic operator induced by $\dagger$, $M = \langle W, \leq, v\rangle$ a preference model, and $\varphi \in \mathcal{L}_0(P)$ a propositional formula. Also given $M_{\star\varphi}=\star(M,\varphi) = \langle W, \leq_{\star\varphi},v\rangle$, take $G = \langle \Phi, \prec\rangle$ be a P-graph inducing $M$ s.t. $\dagger(G,\varphi) = \langle \Phi_\dagger, \prec_\dagger \rangle$ induces $\star(M,\varphi)$.

Take $w,w'\in W$ s.t. $M,w\vDash \varphi$,  $M,w'\not\vDash \varphi$,  and $w \leq w'$. Take $\xi \in \Phi_\dagger$ s.t. $M,w'\vDash \xi$. There is some $\xi' \in \Phi$ s.t. $\varphi \wedge \xi' \vdash  \xi$, $\neg\varphi \wedge \xi \vdash  \xi'$.  As such, $M,w' \vDash \xi'$.

Since $w\leq w'$, either (i) $M,w\vDash \xi'$ or (ii) there is some $\psi' \prec \xi'$ s.t. $M,w\vDash \psi'$ s.t. $M,w'\not\vDash\psi'$. From (ii), we conclude that there is some $\psi\in \Phi_\dagger$ s.t. $\psi\prec \xi$, $\varphi \wedge \psi' \vdash  \psi$, and $\neg\varphi \wedge \psi \vdash  \psi'$.  Since $\neg\varphi \wedge \psi \vdash  \psi'$ and $M,w'\not\vDash \psi'$, we conclude that $M,w'\not\vDash \psi$ and, thus, $M_{\star\varphi},w'\not\vDash \psi$, as $\psi$ is a propositional formula. From (i) and (ii) we conclude that $w\leq_{\star\varphi}w'$.
\end{proof}

We  know that the operation of lexicographic revision \cite{nayak2003dynamic} can be defined by means of transformations on P-graphs \cite{liu2011reasoning}. Also, Nayak et al.~\citey{nayak2003dynamic} have shown that the operation of lexicographic revision is completely characterized by the postulates \textsc{DP}-1, \textsc{DP}-2, and the following postulate known as Recalcitrance (\textsc{Rec}).

\begin{enumerate}
\item[] \textbf{\textsc{(Rec)}} If $w \in \llbracket \varphi \rrbracket$ and $w'\not \in \llbracket \varphi\rrbracket$, then $w <_{*\varphi} w'$.
\end{enumerate}

As such, it is expected that we can characterise \textsc{Rec} as well. Postulate \textsc{Rec} requires for any world satisfying $\varphi$ to be preferred to those not satisfying it. This requirement can be easily guaranteed if all the minimal elements of the changed P-graph which express relevant information (i.e., not equivalent to $\bot$ nor $\top$) imply $\varphi$.   

\begin{prop}\label{prop:Rec}
Let  $\dagger : \mathbb{G}(P)\times \mathcal{L}_0(P) \rightarrow \mathbb{G}(P)$ be a relevant P-graph transformation. If, for any P-graph $G = \langle \Phi, \prec\rangle$ and propositional formula $\varphi \in \mathcal{L}_0(P)$, the P-graph $\dagger(G,\varphi)= \langle \Phi_\dagger, \prec_\dagger\rangle$ satisfies the condition below, then $\dagger$ satisfies \mbox{\textsc{Rec}}:
\begin{itemize}
\item For all $\xi \in \Phi_\dagger$, either $\xi\equiv \top$, $\xi\equiv \bot$, $\xi \vdash \varphi$ or there is some $\psi\in\Phi_\dagger$ s.t. $\psi\prec_\dagger \xi$, $\psi\not\equiv \bot$ and $\psi\vdash \varphi$;
\item There is some $\xi \in \Phi$ s.t. $\xi\vdash \varphi$. 
\end{itemize}
\end{prop}
\begin{proof}[Proof]
It is immediate that if a graph transformation satisfies the condition above, the induced dynamic operator must satisfy \textsc{Rec} since any world satisfying $\varphi$ in a model induced by such a graph would be preferred over any world not satisfying $\varphi$.
\end{proof}

Since \textsc{DP}-3  and \textsc{DP}-4  can be characterised by means of transformations on P-graphs, it is expected that a related postulate might be as well. The postulate of Independence below, proposed by Jin and Thielscher~\citey{JIN06} and independently by \citeonline{booth:jair06}, states that a revision operation may not create arbitrary conditional beliefs in the agent's belief state.

\begin{enumerate}
\item[]\textbf{\textsc{(Ind)}} If $w \! \in \! \llbracket \varphi\rrbracket$ and $w' \! \not \in \! \llbracket \varphi\rrbracket$, then $w \! \leq \! w'$ $\Rightarrow$ $w <_{*\varphi} w'$.
\end{enumerate}

The postulate \textsc{Ind} is, in fact, a stronger form of both \mbox{\textsc{DP}-3} and \textsc{DP}-4. As such, we can provide the following characterisation for it, based on the characterisation of \textsc{DP}-3 and \textsc{DP}-4.

\begin{prop}\label{prop:Ind}
Let  $\dagger : \mathbb{G}(P)\times \mathcal{L}_0(P) \rightarrow \mathbb{G}(P)$ be a relevant P-graph transformation. If, for any P-graph $G = \langle \Phi, \prec\rangle$ and propositional formula $\varphi \in \mathcal{L}_0(P)$, the P-graph $\dagger(G,\varphi)= \langle \Phi_\dagger, \prec_\dagger\rangle$ satisfies the condition below, then $\dagger$ satisfies \mbox{\textsc{Ind}}:
\begin{itemize}
\item For all $\xi' \in \Phi_\dagger$, $\xi'\equiv \varphi$ or there is some $\xi \in \Phi$ s.t. 
\begin{itemize}
\item[(a)] $\varphi \wedge \xi \vdash  \xi'$, 
\item[(b)] $\neg\varphi \wedge \xi' \vdash  \xi$, and
\item[(c)] $\forall \psi' \in \Phi_\dagger$, if $\psi' \prec_\dagger \xi'$ then there is $\psi \in \Phi$ s.t. $\varphi \wedge \psi \vdash  \psi'$, $\neg\varphi \wedge \psi' \vdash  \psi$ and $\psi\prec \xi$.
\item[(d)] if $\xi' \not\vdash \varphi$, there is some $\psi'\prec_\dagger \xi'$ s.t. $\psi'\equiv \varphi$
\end{itemize}
\end{itemize}
\end{prop}
\begin{proof}[Proof]
Let $\dagger$ be a P-graph transformation satisfying the conditions above, $\star$ a dynamic operator induced by $\dagger$, $M = \langle W, \leq, v\rangle$ a preference model and $\varphi \in \mathcal{L}_0(P)$ a propositional formula. Also given $M_{\star\varphi}=\star(M,\varphi) = \langle W, \leq_{\star\varphi},v\rangle$, take $G = \langle \Phi, \prec\rangle$ be a P-graph inducing $M$ s.t. $\dagger(G,\varphi) = \langle \Phi_\dagger, \prec_\dagger \rangle$ induces $\star(M,\varphi)$.

Take $w,w'\in W$ s.t. $M,w\vDash \varphi$, $M,w'\not\vDash \varphi$ and $w \leq w'$. Take $\xi \in \Phi_\dagger$ s.t. $M,w'\vDash \xi$. There is some $\xi' \in \Phi$ s.t. $\varphi \wedge \xi' \vdash  \xi$, $\neg\varphi \wedge \xi \vdash  \xi'$.  As such, $M,w' \vDash \xi'$.

Since $w\leq w'$, either (i) $M,w\vDash \xi'$ or (ii) there is some $\psi' \prec \xi'$ s.t. $M,w\vDash \psi'$ and $M,w'\not\vDash\psi'$. As such, there is $\psi\in \Phi_\dagger$ s.t. $\psi\prec \xi$,  $\varphi \wedge \psi' \vdash  \psi$, and $\neg\varphi \wedge \psi \vdash  \psi'$. Since $\neg\varphi \wedge \psi \vdash  \psi'$ and $M,w'\not\vDash \psi'$, we conclude that $M,w'\not\vDash \psi$ and, thus, $M_{\star\varphi},w'\not\vDash \psi$, as $\psi$ is a propositional formula. Further, since $M,w'\vDash \xi'$, then $\xi'\not\vdash \varphi$. As such, there is some $\psi'\equiv \varphi$. We conclude that (iii) $M,w'\not\vDash \psi'$ and $M,w\vDash \psi'$. From (i), (ii), and (iii), we conclude that $w<_{\star\varphi}w'$.
\end{proof}

%%%%%%%%%%%%%%%%%%%%%%%%%%%%%%%%%%
\section{Deriving P-Graph Transformations from Belief Revision Policies}

The results obtained above can be used to analyse any syntax-based belief revision  policy (P-graph transformations) and derive which belief revision properties (or postulates) it satisfies. On the other hand, these results can also be used to derive implementations for a belief revision operator based on P-graphs. To illustrate this last point, let us examine the case of Lexicographic Revision. 
\begin{defn}
\label{def:RU} \cite{girard2008modal}
Let $M = \langle W, \leq, v\rangle$ be a preference model and $\varphi$ a formula of $\mathcal{L}_0(P)$. We say the preference model  $M_{\Uparrow\varphi} = \langle W, \leq_{\Uparrow\varphi},  v\rangle$ is the result of the lexicographic revision of $M$ by $\varphi$,  where 
$$ w \leq_{\Uparrow \varphi} w' \mbox{ iff } \begin{cases} 
w \leq w' & \mbox{if }w,w' \in \llbracket \varphi \rrbracket\mbox{ or}\\
w \leq w' & \mbox{if }w,w' \not\in \llbracket \varphi \rrbracket\mbox{ or}\\
True & \mbox{if }w\in \llbracket \varphi \rrbracket  \mbox{ and }w' \not\in \llbracket \varphi \rrbracket
\end{cases}$$
\end{defn}

The operation above consists of making each world satisfying $\varphi$ to be strictly more preferable than those not satisfying it, while maintaining the order otherwise. 

It is well-known \cite{nayak2003dynamic} that lexicographic revision is completely characterized by the postulates  (DP-1), (DP-2) and (\textsc{Rec}). Hence, using Propositions~\ref{prop:DP1}, \ref{prop:DP2} and \ref{prop:Rec}, we can construct a P-graph transformation that satisfies these postulates.

A simple P-graph transformation that does satisfy Propositions~\ref{prop:DP1}, \ref{prop:DP2}, and \ref{prop:Rec} is \textit{prefixing} which was proposed by \citeonline{pref} based on the work of \citeonline{andreka2002operators}.

\begin{defn}\label{def:graphprefix}
The prefixing of a P-graph \mbox{$G=\langle \Phi, \prec\rangle$}  by a propositional formula $\varphi \in \mathcal{L}_0(P)$  is the P-graph \mbox{$;(G, \varphi) = \langle \Phi\cup \{\varphi\}, \prec_{;\varphi}\rangle$}, usually denoted by $\varphi;G$ ,  where
\[
\prec_{;\varphi} ~=~ \prec \cup~\{\langle \varphi, \psi\rangle ~|~\psi \in \Phi\}
\]
\end{defn}

Observe that the resulting P-graph maintains all formulas of $\Phi$, thus satisfying Propositions~\ref{prop:DP1} and \ref{prop:DP2}, and includes a formula $\varphi$ (or equivalent to it) that is preferred to all formulas in $\Phi$, thus satisfying Proposition~\ref{prop:Rec}.  As such, the dynamic operator induced by P-graph prefixing satisfies (DP-1), (DP-2) and (\textsc{Rec}).  Since these three postulates completely characterize lexicographic revision, we conclude the following.

\begin{corolary}\label{teo:RU}
Let $M$ be a preference model induced by a P-graph $G$ and $\varphi$ a propositional formula. The model $M_{\Uparrow\varphi}$ is induced by the P-graph $\varphi;G$.
\end{corolary}

%%%%%%%%%%%%%%%%%%%%%%%%%%%%%%%%%%
\section{Negative Results}

While the previous results are encouraging, Souza et al.~\citey{souzakr} already showed that some belief change operators cannot be defined with P-graphs. As such, it must be the case that some postulates in the area cannot be represented by means of transformations on P-graphs - or at least not in a way in which it is jointly consistent with other postulates. To prove such a result, those authors show a simple fact about priority graphs: they cannot encode all the information about the models they induce.

\begin{fact}\label{fact:min}\cite{souzakr}
Let $G = \langle \Phi, \prec\rangle$ a P-graph and $\varphi$ a propositional formula. There is no propositional formula $\mu_\varphi$ s.t. for every model $M = \langle W, \leq_G, v \rangle$ induced by $G$ and all $w\in W$, $w\vDash\mu_\varphi$ iff $w \in \mathit{Min}_{\leq_G} \llbracket \varphi \rrbracket$.
\end{fact} 

Fact~\ref{fact:min} above provides us with some clues to investigate which postulates cannot be characterised through transformations on priority graphs, namely those that refer directly to the minimal worlds of a model. 

One trivial example of such a postulate is the property of an iterated belief change operator to be faithful to AGM's postulates \cite{AGM}, known as postulate \textsc{Faith} below.

\begin{itemize}
\item[] \textbf{(\textsc{Faith})} If $\llbracket \varphi\rrbracket \! \neq \! \emptyset$ then $w\! \in \! \mathit{Min}_{\leq} \llbracket \varphi\rrbracket$ iff $w \! \in \mathit{Min}_{\leq_{*\varphi}} \! W$
\end{itemize} 

Notice that while \textsc{Faith} says something about the minimal worlds of a model, it does not characterise this set in any way. To illustrate it, it suffices to realise that lexicographic contraction satisfies \textsc{Faith} - which describes the change in the agent's belief state by changing the preference of all the worlds satisfying a certain propositional formula $\varphi$. As such, if a P-graph transformation satisfies the postulates \textsc{DP-1} and \textsc{REC} then it satisfies the postulate \textsc{FAITH}. 

Let us then consider some belief change operators requiring a characterisation of the changes in the belief state which is completely dependent on some set of minimal worlds.  To construct such an operator, let us examine the postulate of Conditional Belief Change Minimisation (\textsc{CB}), proposed by Boutilier~\citey{boutilier1993revision}. This postulate states that any iterated belief revision operation must minimise changes of conditional beliefs in the belief state of the agent. 
\begin{enumerate}
\item[]\textbf{(\textsc{CB})} If $w,w' \not\in \mathit{Min}_\leq \llbracket \varphi\rrbracket$, then $w\leq w'$ iff $w \leq_{*\varphi} w'$.
\end{enumerate}

Together with \textsc{Faith}, postulate \textsc{CB} characterises a belief change operator that is completely defined by the changes in the minimal worlds satisfying some formula $\varphi$. As such, it is fairly easy to see that  no graph transformation  satisfies both \textsc{Faith} and \textsc{CB}.

\begin{fact}\label{fact:CB}
No relevant P-graph transformation $\dagger: \mathbb{G}(P)\times \mathcal{L}_0(P) \rightarrow \mathbb{G}(P)$ satisfies both \textsc{Faith} and \textsc{CB}.
\end{fact}
\begin{proof}[Proof]
We suppose  that there is a relevant graph transformation $\dagger: \mathbb{G}(P)\times \mathcal{L}_0(P) \rightarrow \mathbb{G}(P)$ satisfying both \textsc{Faith} and \textsc{CB} and we will derive a contradiction.  Take the preference model $M = \langle \{w_1,w_2,w_3\},\leq_1, v\rangle$ s.t. $w_1<_1w_2<_1w_3$, $M,w_1\vDash \neg p \wedge q $, $M,w_2\vDash p \wedge \neg q$ and ${M,w_3\vDash p \wedge q}$. 
Let $G$ be a P-graph that induces $M$. Since $\dagger$ satisfies both \textsc{Faith} and \textsc{CB}, any dynamic operator $\star$ induced by $\dagger$ must satisfy that $\star(M,p) = \langle \{w_1,w_2,w_3\},\leq_1', v\rangle$ s.t. $w_2<_1'w_1<_1'w_3$ is induced by $\dagger(G,p) = \langle \Phi', \prec'\rangle$, i.e., for any $\xi \in \Phi'$ s.t. $M,w_3\vDash \xi$ either $M,w_1\vDash \xi$ or there is $\psi \in \Phi'$ s.t. $\psi\prec'\xi$, $M,w_1\vDash \psi$ and $M,w_3\not\vDash \psi$ and there is some $\xi \in \Phi'$ s.t. $M,w_1\vDash \xi$ and $M,w_3\not\vDash \xi$.

Consider now the model $M' = \langle \{w_1,w_3\},\leq_2, v\rangle$ s.t. ${w_1 <_2 w_3}$ and $v$ is the same as before. Clearly, $M'$ is induced by $G$ as well. Since $\dagger$ satisfies both \textsc{Faith} and \textsc{CB}, any dynamic operator $\star$ induced by $\dagger$ must satisfy that $\star(M',p) = \langle \{w_1,w_3\},\leq_2', v\rangle$ s.t. $w_3<_2'w_1$ is induced by $\dagger(G,p) = \langle \Phi', \prec'\rangle$. As such, for any $\xi \in \Phi'$ s.t. $M,w_1\vDash \xi$ either $M,w_3\vDash \xi$ or there is $\psi \in \Phi'$ s.t. $\psi\prec'\xi$, $M,w_3\vDash \psi$ and $M,w_1\not\vDash \psi$ and there is some $\xi \in \Phi'$ s.t. $M,w_3\vDash \xi$ and $M,w_1 \not\vDash \xi$. But this is a contradiction with the previous statement, since the valuation $v$ is the same in all models. 
\end{proof}

It is well known, however, that Natural Revision - an iterated revision operation proposed by \citeonline{boutilier1993revision} - satisfies both \textsc{Faith} and \textsc{CB} and is definable on preference models. 

\begin{defn}
\label{def:NatRev}
Let $M = \langle W, \leq, v\rangle$ be a preference model and $\varphi$ a formula of $\mathcal{L}_0(P)$. We say the preference model  $M_{\uparrow\varphi} = \langle W, \leq_{\uparrow\varphi},  v\rangle$ is the result of the natural revision of $M$ by $\varphi$,  where 
$$ w \leq_{\uparrow \varphi} w' \mbox{ iff } \begin{cases} 
w \in \mathit{Min}_\leq \llbracket \varphi\rrbracket\mbox{, or}\\
w \leq w' \mbox{ and }w,w' \not\in \mathit{Min}_\leq \llbracket \varphi\rrbracket
\end{cases}$$
\end{defn}

As such, we can conclude that Natural Revision cannot be represented as a P-graph transformation.

\begin{corolary}\label{cor:CB}
There is no P-graph transformation that induces Natural Revision.
\end{corolary}

This result  shows that some important belief change operations are not definable through P-graph transformations. Notice that previous examples, provided by Souza et al~\citey{souzakr}, have all been contraction operations. What these belief change operations have in common is that their definition is intrinsically characterised by the minimal worlds satisfying a certain formula $\varphi$. In other words, these operations are only well-defined on preference models if we require that preference models be well-founded, a requirement made in Souza's~\citey{souzaphd} definition for these models but not in Girard's~\citey{girard2008modal}. As such, this result reinforces  our intuition that Fact~\ref{fact:min} is the cause of the lack of expressibility of P-graph transformations. 

Notice that, as \textsc{Faith}, postulate \textsc{CB} is not solely responsible for the impossibility of expressing Natural Revision through  P-graph transformations. There is, in fact, a trivial belief change operation that satisfies \textsc{CB} and is expressible by P-graph transformation: the \textit{null change} operation. 

\begin{defn}
Let $M = \langle W, \leq, v\rangle$ be a preference model and $\varphi$ a formula of $\mathcal{L}_0(P)$. We say the preference model  $M_{\circ\varphi} = \langle W, \leq,  v\rangle$ is the result of the null change of $M$ by $\varphi$
\end{defn}

The null operation is the operation of not changing anything in the agent's belief state. It clearly satisfies postulate \textsc{CB} and it is trivially induced by the \textit{null change} P-graph transformation.

\begin{defn}
Let $G \in \mathbb{G}(P)$ be a P-graph and $\varphi \in \mathcal{L}_0(P)$ a propositional formula. We define the \textit{null change} transformation of $G$ by $\varphi$ as $\odot (G,\varphi) = G$. 
\end{defn}

Clearly the only dynamic operator induced by the \textit{null change} transformation  $\odot$ is the \textit{null change} operator $\circ$. As such, $\odot$ satisfies postulate \textsc{CB}.

%%%%%%%%%%%%%%%%%%%%%%%
\section{Related Work}

The AGM approach and the vast literature based on it relies mainly on extralogical characterisation of belief change operations. The first attempt to integrate belief change operation within a logic that we are aware of is the work of \citeonline{segerberg}, which defines Dynamic Doxastic Logic (DDL).

Similar work has focused on embedding specific belief change operations within various epistemic logics to analyse dynamic phenomena in Formal Epistemology  \cite{van2007dynamic,BAL08}. Particularly, \citeonline{girard2008modal} and \citeonline{pref} propose Dynamic Preference Logic (DPL), a dynamic logic which has been used to generalise AGM-like belief change operations \cite{liu2011reasoning,souzakr}.  Aiming to strengthen the connection between DPL and Belief Change, \citeonline{souza:dali} study how well-known belief change postulates can be characterised using DPL axioms. 

While these studies connect Belief Change with Epistemic Logic and provide ways to use the results from one area within the other, their approach is mainly semantic. Research on  Belief Base Change, however, stemming from the work of Hansson~\citey{Hansson}, focus on constructing belief change operators based on syntactic representations of the agent's belief state.

Searching for rich syntactic representations of agents' explicit beliefs, several authors such as Williams~\citey{WIL94,williams1995iterated}, Rott~\citey{rott1991two,rott:shift} and \citeonline{benferhat2002practical} propose different belief base change operations.  These works, however, do not explore how the belief change operations constructed over these syntactic representations are connected to the postulates of Belief Change. 

The work closest to ours is that of \citeonline{delgrande2006iterated} which  considers the notion of iterated belief revision, as studied by \citeonline{darwiche} and \citeonline{nayak2003dynamic}, as a special case of the belief change operation of \textit{merging}. The authors use syntactic structures, similar to prioritized bases, to construct merging operations and show that they satisfy well-known iterated belief revision postulates. More so, the authors propose codifications of these postulates using the syntactic structures proposed by their work, differently than previous work. 

The main drawback of their codification of postulates, in our opinion, is that they are not general enough. The proposed codifications of the postulates are obtained by translating the desired postulates, e.g., Darwiche and Pearl's \citey{darwiche} iterated revision postulates, using the operation of graph prefixing to stand for revision. However, as we know, graph prefixing does not equate revision but represents a specific iterated revision policy known as lexicographic revision. As such, a more general codification of these postulates by means of syntactic representations of the agent's belief state is still an open problem.

\citeonline{liu2011reasoning} has shown that preference models can be encoded by syntactic structures known as P-graphs. Since preference models have been used to model an agent's belief state \cite{girard2008modal,girard2014belief}, Liu's priority graphs can be seen as a syntactic representation of an agent's belief state as well. More yet, \citeonline{liu:deontics} and \citeonline{souzaphd} have shown that this representation can be used to construct well-known belief changing operations from iterated belief change literature. As before, however, the authors do not consider how the formal properties of a belief change operator are reflected in its construction based on transformations of priority graphs.

%%%%%%%%%%%%%%%%%%%%%%%%
\section{Conclusion}

This work has explored codifications of iterated belief change postulates in Dynamic Preference Logic using the syntactic representation of preference models by means of Liu's  priority structures~\cite{liu2011reasoning} known as P-graphs. We provided conditions on P-graph transformations that enforce adherence to belief change postulates of the induced dynamic operators.

Our work can be seen as a generalisation of previous work on the integration of Belief Revision Theory and Dynamic Epistemic Logics that allows the use of the DEL framework to reason about classes of belief change operators. In some sense, this work is the complement of the characterisation of iterated belief change postulates using the proof theory of Dynamic Preference Logic \cite{souza:dali}. We point out that, since priority graphs can be seen as a syntactic form of representing evidence, our work can also be connected to the work on Evidence Logics and Explicit Knowledge \cite{baltag2016beliefs,lorinibases}.

 As illustrated in  Section \textbf{Deriving P-Graph Transformations from Belief Revision Policies}, the characterisations proposed provide a road-map to implement belief change policies in computational systems. Besides, in Section \textbf{Negative Results}  we show that a well-known iterated revision operator cannot be encoded employing graph transformations, and we indicate which policies cannot be implemented as syntactic transformations in the general case of using preference models to reason about belief change.

%%%%%%%%%%%%%%%%%%%%%
 \section{Acknowledgements} 

This study was financed in part by the \textit{Coordena\c{c}\~ao de Aperfei\c{c}oamento de Pessoal de N\'ivel Superior - Brasil (CAPES)} - Finance Code 001. 
 
%%%%%%%%%%%%%%%%%%%%%
\fontsize{9.5pt}{10.5pt}
\selectfont

\end{document}